\documentclass[twocolumn,superscriptaddress,showpacs,amsmath,amssymb,pra]{revtex4}

\usepackage{graphicx}
\usepackage{dcolumn}
\usepackage{bm}
\usepackage{mathrsfs}
\usepackage{amssymb}
\usepackage{amsmath}
\newcommand{\ave}[1]{\langle #1 \rangle}
\newcommand{\bra}[1]{\langle #1|}
\newcommand{\ket}[1]{| #1 \rangle }
\newcommand{\ip}[2]{{\langle #1|}{ #2 \rangle }}

\begin{document}

\title{Towards higher precision and operational use of optical homodyne tomograms}

\author{M. Bellini}
\affiliation{Istituto Nazionale di Ottica, INO-CNR, L.go E. Fermi,
6, I-50125, Florence, Italy}
\affiliation{LENS, Via Nello Carrara
1, I-50019 Sesto Fiorentino, Florence, Italy}
\author{A.S. Coelho}
\affiliation{Instituto de F\'{i}sica, Universidade de S\~{a}o
Paulo, 05315-970 S\~{a}o Paulo, Brazil}
\author{S.N. Filippov}
\email{sergey.filippov@phystech.edu}
\affiliation{Moscow Institute
of Physics and Technology, Moscow Region 141700, Russia}
\author{V.I. Man'ko}
\affiliation{Moscow Institute of Physics and Technology, Moscow
Region 141700, Russia}
\affiliation{P.~N.~Lebedev Physical
Institute, Russian Academy of Sciences, Moscow 119991, Russia}
\author{A. Zavatta}
\affiliation{Istituto Nazionale di Ottica, INO-CNR, L.go E. Fermi,
6, I-50125, Florence, Italy}
\affiliation{LENS, Via Nello Carrara
1, I-50019 Sesto Fiorentino, Florence, Italy}

%

\begin{abstract}
We present the results of an operational use of experimentally
measured optical tomograms to determine state characteristics
(purity) avoiding any reconstruction of quasiprobabilities. We
also develop a natural way how to estimate the errors (including
both statistical and systematic ones) by an analysis of the
experimental data themselves. Precision of the experiment can be
increased by postselecting the data with minimal (systematic)
errors. We demonstrate those techniques by considering coherent
and photon-added coherent states measured via the time-domain
improved homodyne detection. The operational use and precision of
the data allowed us to check for the first time purity-dependent
uncertainty relations and uncertainty relations for Shannon and
R\'{e}nyi entropies.
\end{abstract}


\maketitle


\section{\label{section:introduction} Introduction}

A measurement plays a vital role in the study of quantum physics.
Optical homodyne tomography is merely one among a variety of
measurement techniques, however, its importance and effectiveness
can scarcely be overestimated. The conventional optical homodyne
tomography of one-mode continuous-variable states takes its origin
from the papers~\cite{vogel-risken-1989,raymer,mlynek} and is
instructively described in a series of books and reviews (see,
e.g.,~\cite{leonhardt-book,bachor-book,lvovsky}). The original
goal of the optical homodyne tomography was to infer the quantum
state of light identified with the density operator $\hat{\rho}$
or the Wigner function $W(q,p)$~\cite{wigner}, say. In fact, any
faithfully reconstructed quasi-probability contains a complete
information about the state and can then be used to calculate any
characteristics of the state, for example, its purity.
Unfortunately, no reconstruction procedure is perfect and, what is
more unpleasant, the original errors of experimental data can grow
during the reconstruction. It is generally accepted that the
higher precision of the measurement, the more comprehensive
information is provided and the more sophisticated phenomena can
be observed. The precision is thought to be increased merely by
increasing the number of experimental runs (enlarging an ensemble
of identically prepared states). Clearly, such an approach leads
to a reduction of statistical errors but can hardly cope with
systematic ones (related with the experiment itself). On the other
hand, quantum tomography is a quantitative technique only if we
can evaluate the overall errors presented in the experimental
data. The previous approaches do not give a direct solution of
this problem: the pattern-function reconstruction can provide the
statistical errors only, whereas the maximal likelihood approach
to evaluation of the errors resorts to a bootstrap method whose
result cannot be totally relied on and is
time-consuming~\cite{lvovsky}. In this paper, we propose and apply
in practice a straightforward method to estimate both statistical
and systematic errors, thus, making a step toward a higher
precision of homodyne experiments.

Our approach is based on the initial experimental data and
corresponding histograms of quadrature values. The histograms are
nothing else but an estimation of the quantum tomogram
$w(X,\theta)$. Being a measurable characteristic of the state and
describing the quantum state thoroughly, the tomogram is not only
a powerful tool to reconstruct quasi-distributions but can be used
solely on an equal footing (see the tomographic-probability
representation of quantum
mechanics~\cite{mancini-manko-tombesi-1996,ibort-2009}). Moreover,
as we show in this paper, the tomographic approach can be used to
estimate the errors of histograms and, what is more important, to
calculate \emph{directly} state characteristics (e.g., the purity)
and their errors.

Although different kinds of photon states can be analyzed by the
optical homodyne tomography, we focus our attention on
photon-added
states~\cite{agarwal,dodonov-manko-kor-mukh,dodonov-2002}, whose
experimental
detection~\cite{zavatta-science-2004,zavatta-pra-2005,zavatta-josab-2002,zavatta-LPL-2006,parigi-jpa-2009}
and nonclassical behavior~\cite{zavatta-pra-2007,kiesel-2011} were
demonstrated recently. Moreover, the advanced techniques of photon
addition and photon subtraction enabled us to perform a direct
probe of the commutation relation between photon creation and
annihilation
operators~\cite{parigi-science-2007,kim-jeong-zavatta-prl-2008,zavatta-prl-2009}
as well as accomplish a noiseless
amplification~\cite{zavatta-natphot-2011}.

We use coherent (classical-like) and single-photon added coherent
(non-classical) states to achieve another goal of our paper,
namely, to analyze the accuracy with which the known so far
fundamental quantum relations are fulfilled. Such relations
include, for example, the Heisenberg inequality~\cite{heisenberg}
and its purity-dependent version~\cite{dodonov-manko-1989} as well
as the state-extended uncertainty
relations~\cite{trifonov-2000,trifonov-2002} and the uncertainty
relations for Shannon and R\'{e}nyi
entropies~\cite{hirschman,bialynicki-birula-1975,bialynicki-birula-2006}.
It was shown theoretically in the
papers~\cite{manko-2009,denicola,mmanko-2010,chernega-state-ext-1,chernega-2,mmanko-2011}
how to check all these inequalities by means of the optical
homodyne tomography. In this paper, we present the first
experimental results for some of them. Needless to say that the
accuracy of tomographic data plays the major role in this case.
However, a fulfillment of the Heisenberg uncertainty relation does
not mean that the quantum mechanics in its conventional form is
valid (see,
e.g.,~\cite{narcowich-oconnell,manko-marmo-sudarshan-pla-06}) and
opens a possibility of going beyond the conventional quantum
mechanics (see, e.g.,~\cite{thooft-2003,thooft-2011}). The
violation of the conventional quantum mechanics (if any) could be
detected by the violation of quantum inequalities for highest
moments. In principle, all the highest moments can measured via
homodyne detector as well~\cite{manko-physscr-2010} and the
experimental check of those inequalities is to be discussed
elsewhere.

The paper is organized as follows.

In Sec.~\ref{section:OHT}, the optical homodyne tomography is
shortly reviewed with the emphasis on coherent and photon-added
coherent states. Also, an optimal estimation of the quantum
tomogram is developed and the influence of detection imperfection
is discussed. Sec.~\ref{section:accuracy} is devoted to the
estimation of errors and a brief analysis of the reasons of
systematic errors. In Sec.~\ref{section:operational}, we describe
how to deal with the experimental data in an operational way by
calculating the purity of the state and checking uncertainty
relations mentioned above. Conclusions and prospectives are given
in Sec.~\ref{section:summary}.

\begin{figure}
\includegraphics[width=8cm]{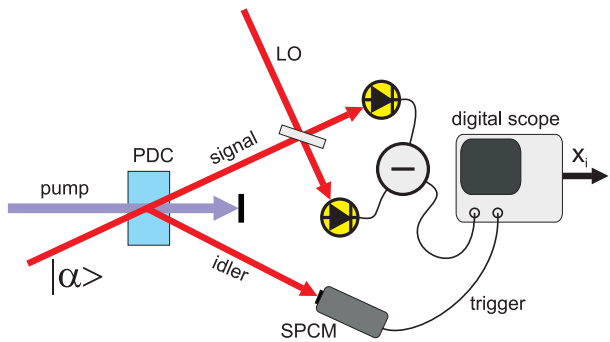}
\caption{\label{fig:setup} (Color online) Schematic of the
experiment for the generation of SPACSs. An UV pulse pumps a
nonlinear crystal to produce stimulated parametric down-conversion
(PDC) in the mode of a seed coherent state. Detection of a single
photon in the conjugated idler mode by a single-photon counting
module (SPCM) heralds the successful generation of a SPACS in the
signal mode, and triggers its homodyne detection. This is
performed by mixing the signal state with a coherent local
oscillator (LO) pulse on a 50$\%$ beam-splitter and measuring the
difference photocurrent produced from two photodiodes at its
outputs.}
\end{figure}


\section{\label{section:OHT}Optical homodyne tomography}
The basic idea of the homodyne tomography is to measure the
quadrature operator $\hat{X}_{\theta} = \hat{Q}\cos\theta +
\hat{P} \sin\theta$, where $\hat{Q}$ and $\hat{P}$ play the role
of position and momentum such that $[\hat{Q},\hat{P}] = i \hbar$
and $\theta \in [0,2\pi]$ is a phase of a strong coherent light
also called the local oscillator (LO). Note that $Q$ and $P$ have
the same units and $\hbar$ is a constant, specified during the
calibration procedure. Fixing the LO phase $\theta$, one can get
access to the probability density distribution (tomogram)
$w(X,\theta) = \langle X_{\theta} | \hat{\rho} | X_{\theta}
\rangle$, where $\hat{X}_{\theta} | X_{\theta} \rangle = X |
X_{\theta} \rangle$. If the tomographic values $w(X,\theta)$ are
specified for all the points $X\in(-\infty,+\infty)$ and
$\theta\in[0,\pi)$, then such an ideal tomogram contains the
complete information about a quantum state.

In this section, we show how to estimate the tomogram of a
coherent state and a single photon added coherent state (SPACS) in
the experiment. Before we move on to the description of the
experiment we briefly discuss the states under investigation.


\subsection{Coherent and SPAC states}
Coherent state $|\alpha\rangle$ is an eigenstate of the photon
annihilation operator $\hat{a}=(\hat{Q}+i\hat{P})/\sqrt{2\hbar}$,
viz., $\hat{a} |\alpha\rangle = \alpha |\alpha\rangle$, where
$\alpha\in\mathbb{C}$. The coherent state $\ket{\alpha}$ is
determined by the following tomogram:
\begin{equation}
\label{tom-coh} w_{\ket{\alpha}}(X,\theta) = \tfrac{1}{\sqrt{\pi
\hbar}} \exp \Big\{ - \Big[\tfrac{X}{\sqrt{\hbar}} - \sqrt{2} (
{\rm Re}\alpha \cos\theta + {\rm Im}\alpha \sin\theta ) \Big]^2
\Big\}.
\end{equation}

The SPACS is defined as $\hat{a}^{\dag} \ket{\alpha} /
\sqrt{1+|\alpha|^2}$ and its tomographic representation reads
(see, e.g.,~\cite{korennoy})
\begin{eqnarray}
\label{tom-spacs} && w_{\hat{a}^{\dag}\ket{\alpha}}(X,\theta) =
\left[ \sqrt{\pi \hbar}
 (1 + |\alpha|^2) \right]^{-1} \nonumber\\
&& \times \Big\{ 2 \left[
\tfrac{X}{\sqrt{\hbar}}-\tfrac{1}{\sqrt{2}}({\rm
Re}\alpha\cos\theta + {\rm Im}\alpha\sin\theta) \right]^2 \nonumber\\
&& \qquad + ({\rm Re}\alpha\sin\theta - {\rm Im}\alpha\cos\theta)^2 \Big\} \nonumber\\
&& \times \exp\Big\{ - \left[ \tfrac{X}{\sqrt{\hbar}} -
\sqrt{2}({\rm Re}\alpha\cos\theta + {\rm Im}\alpha\sin\theta)
\right]^2 \Big\}. \qquad
\end{eqnarray}

\noindent It is not hard to see that in the limit
$|\alpha|\rightarrow \infty$ formula (\ref{tom-spacs}) reduces to
(\ref{tom-coh}), i.e. the SPACS behaves as a coherent state. A
transition from a purely quantum behavior of the SPACS
($\alpha=0$) to a classical-like one ($|\alpha| \gg 1$) was also
observed experimentally~\cite{zavatta-science-2004}.

\begin{figure*}
\includegraphics[width=18cm]{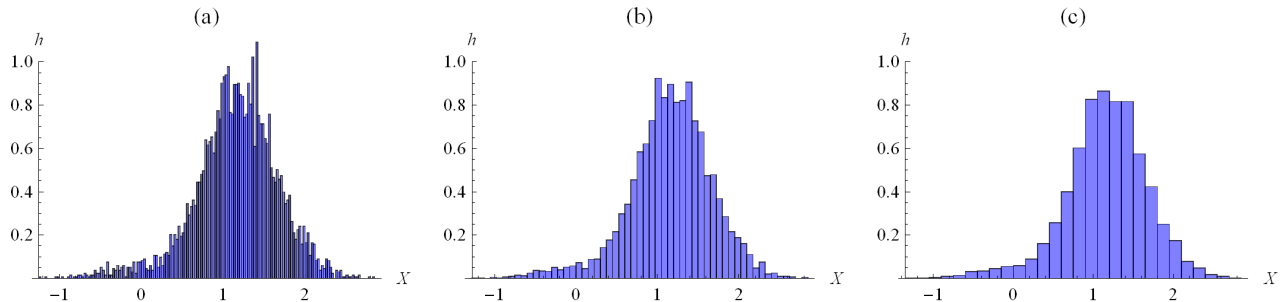}
\caption{\label{bins} (Color online) Histogram $h(X,\pi)$ of a
SPACS for different bin widths: (a) $b=0.025$, great statistical
errors; (b) $b=0.075$; (c) $b=0.15$, larger bin widths can cause
undersampling.}
\end{figure*}


\subsection{Experimental setup}

SPACSs are generated by injecting a coherent state
$|\alpha\rangle$ into the signal mode of an optical parametric
amplifier and exploiting the stimulated emission of a single
down-converted photon into the same mode. Successful SPACS
generation takes place upon detection of a single photon in the
idler mode of the amplifier. Quadrature data are then acquired by
a time-domain balanced homodyne
detector~\cite{zavatta-josab-2002,zavatta-LPL-2006} triggered by
such idler counts. A schematic of the setup, described in detail
in~\cite{zavatta-science-2004,zavatta-pra-2005}, is presented in
Fig.~\ref{fig:setup}.

Acquisition of the quadrature data from the homodyne detector is
accomplished by means of a digital oscilloscope, producing a
sequence of $N=5321$ quadrature values $X$ for each fixed LO
phase. Calibration of $X$-values is accomplished by measuring
vacuum fluctuations when the signal is blocked. In this case,
$\ave{X} =0$ and the variance $\sigma_{XX} = \ave{(X-\ave{X})^2} =
\hbar/2$. Thus, a choice of $\hbar$ is rather arbitrary and we use
$\hbar = \frac{1}{2}$. Once $X$ is calibrated, a state under
investigation is characterized by a collection of points
$\{X_i,\theta_j\}$, where $i=1,\ldots,N$. The phases $\theta_j$
are adjusted by the piezoelectric transducer.


\subsection{Tomogram estimation}

The binned histogram $h(X,\theta_j)$ is known to be constructed
ambiguously because of many possibilities to choose the bin width
$b$. If $b \rightarrow 0$, then the histogram is merely a sum of
delta functions $\delta(X-X_i)$. In fact, for relatively small bin
widths no statistical confidence can be achieved. Conversely, if
$b \rightarrow \infty$, then the histogram transforms into a flat
distribution over the range of $X$, with this uniform distribution
tending to zero. In this case, no useful physical information can
be extracted. Needless to say that none of these two extremal
types of the histogram reflects the behavior of the function
$w(X,\theta)$ predicted by the theory.

Let us now derive an optimal bin width $b$ for purposes of the
optical homodyne tomography. To begin with, the histogram value
$h(X_i,\theta_j)$ at the point $X_i= b i$, $i\in\mathbb{Z}$,
equals $N_i / N b$, where $N_i$ is the number of measured
quadrature values falling into the $i$-th bin $[X_i,X_i + b)$ and
$N$ is the number of all quadrature values. The statistical error
of $h(X_i,\theta_j)$ originates from $N_i$ whose error is
$\sqrt{N_i}$ because the measurement process is assumed to be
Poissonian. For a fixed $N$ we naturally have $N_i \propto b$ (if
$b$ is not too large), then the statistical error of
$h(X_i,\theta_j)$ is $\delta h_{\rm stat} = \sqrt{h(X_i,\theta_j)
/  N b}$. For relatively large bin widths the statistical error is
negligible, however, the effect of undersampling the quadrature
distribution comes into play~\cite{leonhardt-1996}. The main idea
is that the theoretical tomogram $w(X,\theta_j)$ exhibits
oscillating behavior with respect to $X$, with the scale of
oscillations being $\sim \pi / \sqrt{2d}$, where $d$ is the number
of Fock states significantly contributing to the state under
investigation. The experimental histogram $h(X,\theta_j)$ should
reflect those oscillations rather than conceal them. Then, the
error of undersampling for the histogram value $h(X_i,\theta_j)$
can be evaluated as $\delta h_{\rm und} = h(X_i,\theta_j) b
\sqrt{2d} / \pi$. The resulting error $\delta h_{\rm stat} +
\delta h_{\rm und}$ takes minimal value if $b = b_{\rm opt} \equiv
[ \pi / 4\sqrt{2} h(X_i,\theta_j) N d ]^{1/3}$. Note that $b$ has
the same functional dependence $\propto 1/\sqrt[3]{N}$ as the
Scott's choice $b=3.5\sigma/\sqrt[3]{N}$, where $\sigma$ is the
standard deviation of $X$~\cite{scott}. Note also that the optimal
bin width should increase for lower values $h(X_i,\theta_j)$, for
instance at the end of the distribution tails. For practical
purposes the alternating bin widths are, however, not very
convenient since they complicate data processing.

We plot some examples of histograms for different values of the
bin width in Fig.~\ref{bins}. For our further analysis,  we choose
$b=0.075$ which is close to the average optimal value $b_{\rm opt}
\approx 0.06$ (we put $h(X_i,\theta_j) \approx 1/\sqrt{2\pi}$,
$N=5321$, $d \sim 1$, and scale $b_{\rm opt}$ by a factor
$\sqrt{\hbar} = \frac{1}{\sqrt{2}}$). In our case, this bin width
is also close to $(\max X-\min X)/\sqrt{N} \approx 0.055$ known as
the square-root choice. For the normal distribution (coherent
state) the Scott's choice gives $b=0.14$ and the Sturges' formula
results in $\lceil \log_2 N + 1 \rceil \approx 13$ bins ($b=0.3$).
We choose $b=0.075$ to guarantee the statistical confidence and
prevent the data from undersampling. The latter fact is important
to observe the cases when the theoretical function $w(X,\theta)$
tends to zero in the middle of the range of $X$ as it takes place,
e.g., for an ideal SPACS (see formula (\ref{tom-spacs}) with real
$\alpha$ and $\theta=0$ or $\pi$).

\begin{figure*}
\begin{center}
\includegraphics[width=18cm]{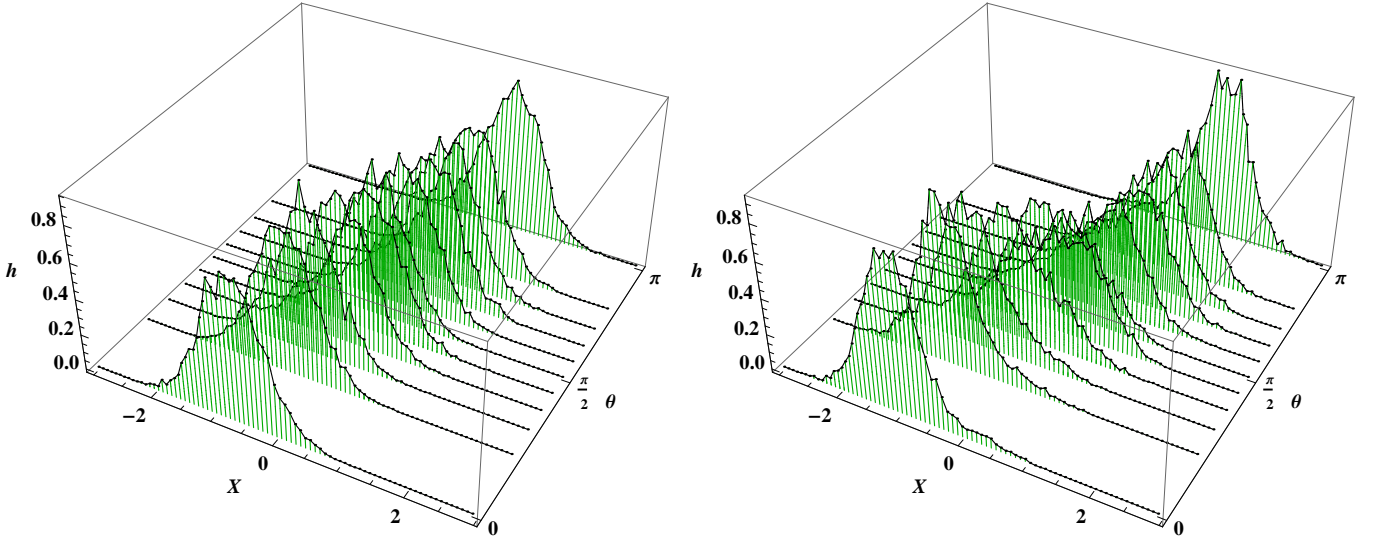}
\caption{\label{tomogram-coh} (Color online) Typical histograms
$h(X,\theta_j)$ of a coherent state (left) and a SPACS (right).}
\end{center}
\end{figure*}

To illustrate the estimated tomograms of a coherent state and a
SPACS, we present a series of historgrams $h(X,\theta_j)$
constructed on the basis of the experimental data with eleven LO
phases $\{\theta_j\}_{j=1}^{11}$ within the region $[0,\pi]$ (see
Fig.~\ref{tomogram-coh}).

\subsection{\label{subsection:detection-imperfection} Detection imperfection}
Let us be reminded that for a real $\alpha$ and the LO phase
$\theta = 0$ the tomogram (\ref{tom-spacs}) of a SPACS is
$w_{\hat{a}^{\dag}\ket{\alpha}}(X,0) \propto
(X/\sqrt{\hbar}-\alpha/\sqrt{2})^2
\exp[-(X/\sqrt{\hbar}-\sqrt{2}\alpha)^2]$ and takes on zero value
if $X=\alpha\sqrt{\hbar/2}$ ($X=-\alpha\sqrt{\hbar/2}$ if
$\theta=\pi$). However, one can hardly observe such property in
Figs.~\ref{bins} and \ref{tomogram-coh}. This is caused by the
fact the detection efficiency $\eta<1$. The detection efficiency
comprises all kinds of losses including the finite efficiency of
photodetectors. Due to the imperfect detection, the measured
histograms are smoothed and there is no zero point
$X=\alpha\sqrt{\hbar/2}$ anymore. In fact, one actually measures
not the prepared state but its convolution with a vacuum (that
impinges a fictitious beamsplitter with transmittivity $\eta$ in
front of an ideal quadrature detector). In terms of the Wigner
fuction, the measurable state $W^{\rm det}(q,p)$ is connected with
the originally prepared state $W(q',p')$ by the following
relation:
\begin{eqnarray}
\label{convolution} W^{\rm det}(q,p) &=& \frac{1}{\pi (1-\eta)}
\iint d q' ~ d p' ~
W(q',p') \nonumber\\
&& \times \exp \left[ - \frac{(q-\sqrt{\eta} q')^2 +
(p-\sqrt{\eta} p')^2}{1-\eta} \right]. \quad
\end{eqnarray}

\noindent It is not hard to see that a coherent state
$\ket{\alpha}$ transforms into the coherent state
$\ket{\sqrt{\eta}\alpha}$ under convolution (\ref{convolution}).
On the other hand, a SPACS remains no longer a SPACS and the
measurable state is given by the following Wigner function:
\begin{eqnarray}
&& W_{\hat{a}^{\dag}\ket{\alpha}}^{\rm det}(q,p) = \left[ \pi(1+|\alpha|^2) \right]^{-1} \nonumber\\
&& \times \Big\{ 1+2\eta \Big[ \Big( q - \tfrac{2\eta -
1}{\sqrt{2\eta}} {\rm Re}\alpha \Big)^2 + \Big( p - \tfrac{2\eta -
1}{\sqrt{2\eta}} {\rm Im}\alpha
\Big)^2 - 1 \Big] \Big\} \nonumber\\
&& \times \exp \left[ - \left( q - \sqrt{2\eta} {\rm Re}\alpha
\right)^2 - \left( p - \sqrt{2\eta} {\rm Im}\alpha \right)^2
\right].
\end{eqnarray}

\noindent Then, for a SPACS with real $\alpha$, the theoretical
prediction of the measurable quadrature distribution is
\begin{eqnarray}
&& \!\!\!\!\!\! w_{\hat{a}^{\dag}\ket{\alpha}}^{\rm det}(X,0) = \left[ \sqrt{\pi \hbar} (1+\alpha^2) \right]^{-1} \nonumber\\
&& \!\!\!\!\!\! \times \Big[ 1 - \eta +2\eta \Big(
\tfrac{X}{\sqrt{\hbar}} - \tfrac{2\eta - 1}{\sqrt{2\eta}} \alpha
\Big)^2 \Big]  \exp \Big[ - \Big( \tfrac{X}{\sqrt{\hbar}} -
\sqrt{2\eta} \alpha \Big)^2 \Big], \nonumber\\
\end{eqnarray}

\noindent which has no zeros and correctly describes the
experimental histograms in Figs.~\ref{bins} and
\ref{tomogram-coh}.

It is worth noting that the purity of the state can reveal the
detection imperfection. Although a coherent state remains pure in
transformation (\ref{convolution}), a SPACS does not. Indeed, the
purity of the detectable SPACS reads
\begin{equation}
\label{purity-spacs} \mu_{\hat{a}^{\dag}\ket{\alpha}}^{\rm det} =
2\pi \iint d q \, d p \left[ W_{\hat{a}^{\dag}\ket{\alpha}}^{\rm
det}(q,p) \right]^2 = 1 - \frac{2 \eta (1-\eta)}{(1+|\alpha|^2)^2}
\end{equation}

\noindent and is less than 1 whenever $0 < \eta < 1$ (if $\eta=0$,
then the vacuum noise is only detected).

In what follows, we will concentrate on the accuracy of the
experimental histograms $h^{\rm det}(X,\theta)$ and theoretical
tomograms $w^{\rm det}(X,\theta)$. Thus, we will operate with the
``detectable" state (not the originally prepared one). Further, we
will omit the superscript $^{\rm det}$ wherever it is clear from
the context. In fact, deconvolution of formula (\ref{convolution})
is known to be difficult to perform with experimentally given
quasiprobabilities~\cite{leonhardt-book} and this is beyond the
scope of present paper. We can refer the interested reader to the
paper~\cite{filippov}, where a similar deconvolution problem is
solved, namely, an extraction of the originally prepared microwave
quantum state from a noisy output of a linear amplifier is
considered.


\section{\label{section:accuracy} Accuracy of optical homodyne tomograms}
Further progress of applied quantum information technologies and
fundamental experiments depends greatly on the accuracy of
measurement data. In optical homodyne detection of radiation
field, one usually restricts oneself by the initial calibration of
the detector outcomes. Namely, blocking photons of the signal mode
results in the vacuum state, whose quadrature distribution is to
be centered at point $X=0$ and have the dispersion $\langle X^2
\rangle = \hbar/2$ for any phase of the local oscillator. However,
in practice, a drift of the scheme parameters or an extra noise
can occur during the experiment. In view of this, for practical
purposes it is extremely important to trace the adequacy of the
data being collected either in real time or during postprocessing.
Also, the method would be beneficial if it were based on the data
themselves without much additional information. In this section,
we present and apply such a method.

A true tomogram $w(X,\theta)$ is known to satisfy the relation
$w(X,\theta) = w(-X, \theta + \pi)$. This fact was previously used
to claim that the quadrature distribution for LO phases $\theta
\in [0,\pi)$ determine a quantum state thoroughly. As a result,
the phases out of this range were disregarded in experiments,
although they naturally provide an efficient way to check the
accuracy of the data. In what follows we show that one can
efficiently use the peculiar property $w(X,\theta) = w(-X, \theta
+ \pi)$ to check whether the data are
adequate~\cite{filippov-fss-11}. Moreover, one can evaluate the
accuracy of the histograms.

For example, an imbalance of the optical scheme or photodetectors'
efficiencies would result in values $X$ shifted by some $x_{\rm
imb}$. In this case, the distributions $w(X,\theta)$ and $w(-X,
\theta + \pi)$ as functions of variable $X$ would be shifted with
respect to each other by the magnitude $2x_{\rm imb}$. In case of
different photodetector efficiencies, $\eta_1$ and $\eta_2$, the
shift $x_{\rm imb} \propto (\eta_1 - \eta_2) I$, where $I$ is the
LO intensity. Analogous mismatch between tomograms can take place
due to a low frequency electronic noise at the input of the
digital scope, the shift alternating in time.

Another reason of possible deviation of $w(X,\theta_1)$ from
$w(-X, \theta_2)$, where $\theta_2$ is supposed to be equal to
$\theta_1+\pi$, can occur due to inaccuracy in the LO phase
control. Especially clearly this type of data mismatch is seen for
a coherent state $|\alpha\rangle$, for which the distribution
$w(X,\theta_1)$ is shifted with respect to the distribution
$w(X,\theta_2)$ by $x_{\delta\theta}= \sqrt{2} \hbar [ {\rm
Re}\alpha (\cos\theta_1+\cos\theta_2) + {\rm Im}\alpha
(\sin\theta_1+\sin\theta_2) ]$ along $X$-axis. For
$\theta_{1,2}=\theta \pm \delta\theta/2$, the shift
$x_{\delta\theta}$ is approximately equal to $\sqrt{2}\hbar({\rm
Re}\alpha \sin\theta - {\rm Im}\alpha \cos\theta) \delta\theta +
\hbar({\rm Re}\alpha \cos\theta + {\rm Im}\alpha \sin\theta)
\delta\theta^2/\sqrt{2}$.

\begin{figure*}
\includegraphics[width=18cm]{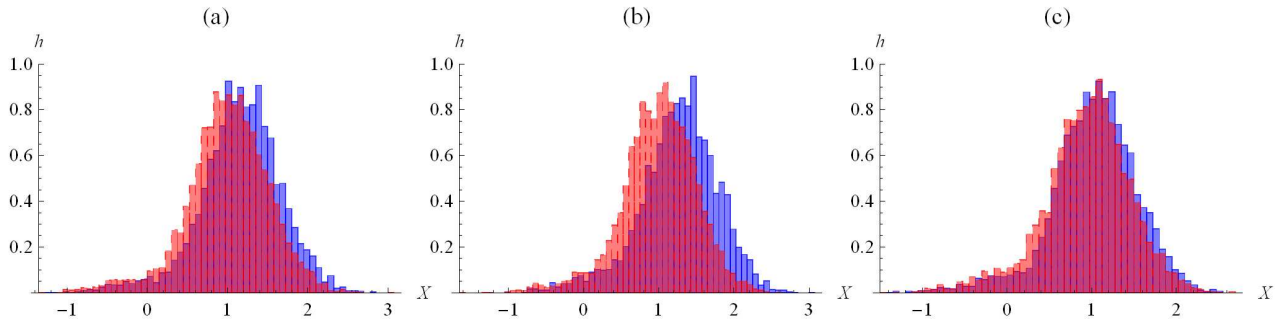}
\caption{\label{checking-h-spacs}  (Color online) Histograms
$h(X,\pi)$ (blue solid lines) and $h(-X,0)$ (red dashed lines) of
SPACS from first column of Table~\ref{table:shifts}: (a) data set
\#1; (b) data set \#2; (c) data set \#3.}
\end{figure*}

In order to demonstrate the method above, we consider a mismatch
between histograms $h(X,\pi)$ and $h(-X,0)$, which should be
coincident according to the theory. Typical histograms of a SPACS
are depicted in Fig.~\ref{checking-h-spacs} for three data sets
corresponding to $\sqrt{\eta}\alpha = 0.64$.

One can readily notice the deviation of histograms $h(X,\pi)$ and
$h(-X,0)$ for data sets \#1 and \#2. The shift between these
histograms is evaluated as the difference between mean values of
the distributions (to be precise, the shift $x = \ave{X_{\pi}} -
\ave{-X_{0}} = \ave{X_{\pi}} + \ave{X_{0}}$). The experimentally
determined shifts of the histogram $h(-X,0)$ with respect to
$h(X,\pi)$ are summarized for coherent states and SPACS states of
different intensities in Table~\ref{table:shifts}.

\begin{table}
\caption{\label{table:shifts} Shifts $x=\ave{X_{\pi}} +
\ave{X_{0}}$ of the histogram $h(-X,0)$ with respect to $h(X,\pi)$
for detected coherent and SPAC states of different intensities.
The amplitude $\sqrt{\eta}\alpha$ of the detected coherent state
is evaluated by the experimentally measured value
$(\ave{X_{\pi}^{\rm coherent}}-\ave{X_{0}^{\rm coherent}})/2$.}
\begin{center}
\begin{tabular}{|c|c||c|c|c|c|}
  \hline
  \multicolumn{2}{|c||}{Data} &   \multicolumn{4}{c|}{Detected amplitude $\sqrt{\eta}\alpha$}
  \\ \cline{3-6}
  \multicolumn{2}{|c||}{set}  & ~~~~0.64~~~~ & ~~~~0.82~~~~ & ~~~~1.25~~~~ & ~~~~1.73~~~~ \\
  \hline\hline
   \#1 & coherent        & 0.14 & 0.15 & $-$0.15 & $-$0.17 \\ \cline{2-6}
       & SPACS           & 0.16 & 0.20 & $-$0.08 & $-$0.10 \\
  \hline
   \#2 & coherent        & 0.26 & $-$0.21 & 0.02 & 0.23 \\ \cline{2-6}
       & SPACS           & 0.26 & $-$0.14 & 0.05 & 0.27 \\
  \hline
   \#3 & coherent        & 0.03 & $-$0.12 & 0.003 & 0.36 \\ \cline{2-6}
       & SPACS           & 0.09 & $-$0.07 & 0.07  & 0.40 \\
  \hline
\end{tabular}
\end{center}
\end{table}

While $|\alpha|$ is getting larger, one expects the error of
fixing the LO phase $\delta\theta$ to get smaller (since the phase
control is based on observing an interference picture which
becomes clearer for larger $|\alpha|$). On the other hand, in case
of real $\alpha$ and LO phase $\theta = 0$, the shift
$x_{\delta\theta}$ equals $\hbar \alpha \delta\theta^2/\sqrt{2}$
and can be non-monotonic with respect to $\alpha$ because of an
additional factor.

Let us now analyze how a mismatch between distributions
$w(X,\theta)$ and $w(-X, \theta + \pi)$ affects the accuracy of
the data and allows evaluating the experimental errors of some
state characteristics.

A natural characteristic, which shows the closeness of two
probability distributions $p_1(X)$ and $p_2(X)$, is the
Bhattacharyya coefficient~\cite{bhattacharyya} defined as $B =
\int \sqrt{ p_1(X) p_2(X) } d X$. The Bhattacharyya coefficient
$B$ equals 1 if and only if distributions $p_1(X)$ and $p_2(X)$
are identical.

Let $\rho_1$ be a state reconstructed from the homodyne tomograms
$w(X,\theta)$, $\theta \in [0,\pi)$, and $\rho_2$ be a state
reconstructed from the tomograms $w(X,\theta)$, $\theta \in
[\pi,2\pi)$. Provided ideal tomograms the states $\rho_1$ and
$\rho_2$ are identical. Experimental data result in two different
states, the fidelity $F(\rho_1,\rho_2)={\rm Tr} \sqrt{
\sqrt{\rho_1} \rho_2 \sqrt{\rho_1} }$ between which indicates the
accuracy of measured data and can be used as an estimate of the
fidelity between the evaluated (reconstructed) state $\rho_{\rm
est}$ and the actual state $\rho$, i.e. $F(\rho_1,\rho_2) \approx
F(\rho,\rho_{\rm est})$. Important for us is the fact that
$F(\rho_1,\rho_2)$ satisfies the following
relation~\cite{filippov-manko-distances}:
\begin{equation}
\label{fidelity-bhattacharyya} F \le \min_{\theta \in [0,\pi]}
\int \sqrt{w(X,\theta) w(-X, \theta + \pi)} d X,
\end{equation}

\noindent that is the fidelity is limited by the minimal
Bhattacharyya coefficient $B_{\theta}$ for the distributions
$p_1(X)=w(X,\theta)$ and $p_2(X)=w(-X, \theta + \pi)$.

In principle, formula (\ref{fidelity-bhattacharyya}) implies
minimization of $B_{\theta}$ over all experimentally accessible LO
phases $\theta$. In this research, we restrict ourselves by an
illustration of the method of fidelity evaluation and present some
values $B_{\theta=0}$ calculated for the data from the first
column of Table~\ref{table:shifts}: 98.70\% and 98.32\%, 96.20\%
and 95.59\%, 99.67\% and 99.26\% for the coherent and SPAC states
from data sets \#1, \#2, and \#3, respectively. Here, we have
calculated the integral (\ref{fidelity-bhattacharyya}) by
replacing $w(X,\theta) \rightarrow h(X,\theta)$ and using the
trapezoid method~\cite{korn}, with the error of calculation being
$-\frac{1}{12} b^3 \frac{d^2}{dX^2} \sqrt{w(X,0) w(-X, \pi)} <
0.004\%$. Once fidelity is evaluated, one can use this knowledge
to evaluate the accuracy of other state characteristics~(see,
e.g.,~\cite{dodonov-jpa-2012,dodonov-jrlr-2011}).

Given tomograms $w(X,\theta)$ for two regions of the LO phases
$\theta \in [0,\pi)$ and $\theta \in [\pi,2\pi)$, it is possible
to evaluate the error of the mean value of any physical quantity
$A$. Indeed, $\Delta A = | {\rm Tr}[(\rho_1 - \rho_2) A] |$, where
$\rho_1$ and $\rho_2$ are defined as above. However, for some
quantities one does not have to reconstruct the states and can use
tomograms directly. For instance, the moment $\langle X_{\theta}^n
\rangle = \int X^n w (X, \theta) d X $ is determined with the
experimental error $\Delta(X_{\theta}^n) = \int X^n | w (X,
\theta) - w (-X, \theta + \pi)| d X$. For example, for the data
set \#3 from the first column of Table~\ref{table:shifts}, the
second moment $\ave{q^2} \equiv \ave{X_{\theta=0}}$ equals
$\langle q^2 \rangle = 0.63 \pm 0.04$ for the coherent state and
$\langle q^2 \rangle = 1.23 \pm 0.16$ for the SPACS. For the first
moments $\langle q \rangle$ the errors are merely the shifts 0.03
and 0.09, respectively. The error bars of those quantities are of
the same order for other LO phases.

To conclude this section, a relatively simple analysis of the
homodyne tomographic data enables one to check their adequacy and
evaluate their accuracy. As a result, one can postselect and use
further only those data that meet the desired accuracy. Moreover,
a mismatch between tomographic data can indicate a reason and
nature of extra noise. The latter fact opens up new vistas of the
optical homodyne tomography in metrology.


\section{\label{section:operational} Operational use of the tomographic data}
In this section, we are going to reveal some relevant information
about a quantum state just using the tomographic data and
circumventing a reconstruction of the density operator or the
Wigner function. Also, we are checking if the data satisfy some
theoretically predicted inequalities. In this section, explicit
numerical values of quantities of interest are calculated for data
set \#3 from the first column of Table~\ref{table:shifts} which
exhibits relatively small systematic errors.


\subsection{\label{section:purity}Purity}
Purity $\mu = {\rm Tr} \rho^2$ is an important state
characteristic which can set some limitations on the use of the
state in applications. A conventional approach to determine the
state purity from the optical tomogram is to reconstruct the
density matrix or the Wigner function via some improved
modifications of the inverse Radon transform~\cite{benichi} or the
maximum likelihood method~\cite{hradil} and then substitute them
in some integral relations to calculate the purity. Recently, the
state purity has been also evaluated from quadratures'
uncertainties~\cite{porzio-solimeno-2011}. This method is easy to
use but the evaluation gives a correct value only for Gaussian
states. Here, we use the tomographic data directly and calculate
the true purity without any intermediate reconstruction of the
density operator or a quasiprobability distribution. Moreover, no
assumption about the state being Gaussian is needed.

The purity is known to be expressed through the optical tomogram
as follows~\cite{omanko-2009}:
\begin{eqnarray}
\label{purity-theor} \mu_{\rm theor} &=& \frac{1}{2\pi}
\int_{0}^{+\infty} d r ~ r \iint_{-\infty}^{+\infty} d X d Y \,
e^{-i(X+Y)r} \nonumber\\
&& \times \int_{0}^{2\pi} d \theta \, w(X,\theta) w(-Y,\theta),
\end{eqnarray}

\noindent where the sequence of taking integrals is chosen for the
easiest data processing. If the tomograms satisfied the relation
$w(X,\theta) = w(-X, \theta+\pi)$, the calculated value of $\mu$
would be real. In fact, one would have
\begin{eqnarray}
&& \int_{0}^{2\pi} d \theta \, w(X,\theta) w(-Y,\theta) \nonumber\\
&& = \int_{0}^{\pi} d \theta \, [ w(X,\theta) w(-Y,\theta) +
w(-X,\theta) w(Y,\theta) ] \qquad
\end{eqnarray}

\noindent and, consequently,
\begin{eqnarray}
\label{purity-cos} \mu &=& \frac{1}{\pi} \int_{0}^{+\infty} d r ~
r \iint_{-\infty}^{+\infty} d X d Y \, \cos[(X+Y)r]
\nonumber\\
&& \times \int_{0}^{\pi} d \theta \, w(X,\theta) w(-Y,\theta).
\end{eqnarray}

The obtained formula is beneficial when the homodyne data are
acquired only for the LO phases in the range $[0,\pi]$ (although
it is impossible to evaluate the accuracy of $\mu$ then). As we
already know, in practice the requirement $w(X,\theta) = w(-X,
\theta+\pi)$ is not precisely met. Then the imaginary part of
expression (\ref{purity-theor}) can serve as the error bar of the
purity. It can be also calculated as follows:
\begin{eqnarray}
\label{delta-mu} && \!\!\!\!\!\!\! \Delta\mu = ({\rm Tr} \rho_1^2
- {\rm Tr}
\rho_2^2)/2 \nonumber\\
&& \!\!\!\!\!\!\! = \frac{1}{2\pi} \int_{0}^{+\infty} r d r
\iint_{-\infty}^{+\infty} d X d Y
\cos[(X+Y)r] \nonumber\\
&& \!\!\!\!\!\!\! \times \int_{0}^{\pi} d \theta [ w(X,\theta)
w(-Y,\theta) - w(X,\theta+\pi) w(-Y,\theta+\pi)]. \nonumber\\
\end{eqnarray}

\begin{figure}
\includegraphics[width=8cm]{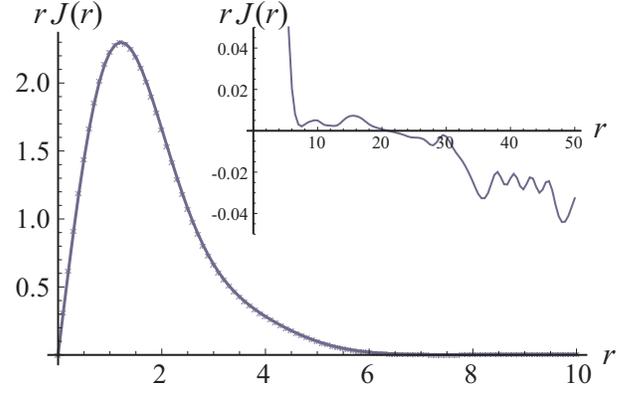}
\caption{\label{fig:purity-calculation} (Color online) Calculation
of purity of the SPACS. The purity is the area under the curve
divided by $2\pi$. Function $J(r)$ deviates from 0 for $r > 8$ due
to calculational errors.}
\end{figure}

Given experimental histograms $h(X,\theta_j)$, we first calculate
the sum $\frac{1}{2}\sum_{j=1}^{N_{\theta}-1} [ h(X_i,\theta_j)
h(-Y_k,\theta_j) + h(X_i,\theta_{j+1}) h(-Y_k,\theta_{j+1}) ]
(\theta_{j+1} - \theta_j)$ for any pair of bin coordinates
$(X_i,Y_k)$, i.e. the evaluation of the function
$P(X,Y)=\int_{0}^{\pi} d \theta w(X,\theta) w(-Y,\theta)$ via the
trapezoid method. The error of this evaluation is roughly equal to
$-\frac{1}{12}(\theta_{j+1} - \theta_j)^3
\frac{\partial^2}{\partial\theta^2}[w(X_i,\theta) w(-Y_k,\theta)]
\lesssim
\frac{2\pi^2|\alpha|^2}{3(N_{\theta}-1)^3}\exp(-X_i^2-Y_k^2)
\lesssim 0.003 \exp(-X_i^2-Y_k^2)$ for the states in question.
Calculation of the integral $J(r) = \iint_{-\infty}^{+\infty} d X
d Y \cos[(X+Y)r] P(X,Y) $ is substituted by the calculation of the
sum $\sum_{X_i,Y_k} b^2 \cos[(X_i+Y_k)r] P(X_i,Y_k)$ for any fixed
$r$. This evaluation contains two types of errors: the first one
originates from the error of the function $P(X_i,Y_k)$ and equals
$0.01e^{-r^2/2}$, and the second one is due to evaluation of the
integral $\iint_{-\infty}^{+\infty} d X d Y$ by the sum
$\sum_{X_i,Y_k}$ and equals $-\frac{1}{12} b^4 \left(
\frac{\partial^2}{\partial X^2} + \frac{\partial^2}{\partial
Y^2}\right) \cos[(X_i+Y_k)r] P(X_i,Y_k) \lesssim 10^{-5} (2+r^2)$.
Evaluation of the function $rJ(r)$ for the SPACS is presented in
Fig.~\ref{fig:purity-calculation}. Deviation of $J(r)$ from 0 for
values $r>8$ is to be assigned to the second type of the error.
Finally, the purity parameter (\ref{purity-cos}) is calculated via
integrating the function $rJ(r)$ in the range $[0,R]$, where the
upper limit $R$ is chosen in such way that the integral
$\mu(R):=\int_{0}^{R}rJ(r)dr$ is saturated and does not depend of
$R$. The error of calculating $\mu(R)$ can be evaluated as
$\Delta_{\rm calc} \mu(R) \lesssim
0.01\int_{0}^{+\infty}re^{-r^2/2}dr + 10^{-5} \int_{0}^{R}
r(2+r^2)dr \approx 0.01 + 3 \cdot 10^{-6} R^4 $. We choose $R=8$
and obtain $\mu = 1.00$ for the coherent state and $\mu = 0.83$
for the SPACS, the error of calculation being $\Delta_{\rm calc}
\mu \lesssim 0.02$. Similarly, we use formula (\ref{delta-mu}) to
calculate the error $\Delta\mu$ originating from the inaccuracy of
the original data. The direct calculation yields 0.035 for the
coherent state and 0.039 for the SPACS, which is slightly greater
than the error of calculation $\Delta_{\rm calc}\mu$. To resume,
we obtain $\mu = 1.00 \pm 0.04$ and $\mu = 0.83 \pm 0.04$ for the
coherent state and the SPACS, respectively (the overall error is
estimated as $[(\Delta_{\rm calc} \mu)^2 + (\Delta\mu)^2]^{1/2}$).
It is worth noting that the obtained purity $\mu = 0.83$ of the
detected SPACS exactly coincides with what is predicted by
Eq.~(\ref{purity-spacs}) if nominal values $\eta=0.6$ and
$\alpha=0.83$ are used.

We note that the errors are easily and naturally estimated in our
approach in contrast to the approaches based on the density
operator reconstruction that involves a rather time-consuming
bootstrap method for evaluation of the errors by the maximal
likelihood technique. Note that a calculation of the purity for a
given density operator also results in additional calculational
errors.

\subsection{Fidelity}

Usually, an experiment is aimed at producing a specific pure
quantum state, $\ket{\psi}$ say. Experimentally determined
quadrature distributions $w_{\rm exp}(X,\theta)$ allow calculating
the fidelity $F^2=\bra{\psi}\rho\ket{\psi}$, where $\rho$ is an
actually detected state. Similarly to formula (\ref{purity-cos}),
we have~\cite{manko-elaf-11}: $\bra{\psi}\rho\ket{\psi} =
\frac{1}{\pi} \int_{0}^{+\infty} r \, dr \iint_{-\infty}^{+\infty}
d X d Y \, \cos[(X+Y)r] \int_{0}^{\pi} d \theta \,
w_{\psi}(X,\theta) w_{\rm exp}(-Y,\theta)$, where the analytical
function $w_{\psi}(X,\theta) = |\ip{X_\theta}{\psi}|^2$ is easily
computed through the desired state $\ket{\psi}$. For instance, if
$\ket{\psi}$ is a superposition of a finite number of Fock states
$\ket{n}$, then an explicit formula for $w_{\psi}(X,\theta)$ is
found, e.g.,~in Ref.~\cite{filippov-fss-11}. Since the function
$w_{\psi}(X,\theta)$ is known precisely, the error of the quantity
$F^2$ equals $\frac{1}{2\pi} \int_{0}^{+\infty} r \, dr
\iint_{-\infty}^{+\infty} d X d Y \cos[(X+Y)r] \int_{0}^{\pi} d
\theta \, w_{\psi}(X,\theta) [w_{\rm exp}(-Y,\theta) - w_{\rm
exp}(Y,\theta+\pi)]$.


\subsection{Experimental check of uncertainty relations}


\subsubsection{Heisenberg inequality}
Since the main difference between two histograms $h(X,\theta)$ and
$h(-X,\theta+\pi)$ is essentially the shift, the variances
$\sigma_{X_{\theta}X_{\theta}} = \langle X_{\theta}^2 \rangle -
\langle X_{\theta} \rangle^2$ differ not so severely as the second
moments. In fact, in our case we have $\Delta \sigma_{qq} = 0.004$
for the coherent state and $0.013$ for the SPACS.

In this subsection, we are going to check if the Heisenberg
uncertainty relation $\sigma_{qq} \sigma_{pp} \ge \hbar^2 /4$
holds true and what is the extent to which it is fulfilled.
Certainly, thanks to the initial calibration of the apparatus by
the vacuum state, we can adjust $\hbar = \frac{1}{2}$ and check
the inequality for any other states but not the vacuum itself. The
errors of determining second moments are evaluated as described in
Sec.~\ref{section:accuracy}.

For the coherent state we have $\sigma_{qq} \sigma_{pp} = 0.0612
\pm 0.0014$, which coincides with $0.0625$ within the error bar.
For the SPACS we obtain $\sigma_{qq} \sigma_{pp} = 0.101 \pm 0.006
> 0.0625$. The coherent state has the minimal uncertainty indeed and, therefore, is pure.
This result is in agreement with the detection imperfection
discussed in Sec.~\ref{subsection:detection-imperfection} because
the imperfect detection results in $|\alpha\rangle \rightarrow
|\sqrt{\eta} \alpha \rangle$, i.e. the pure coherent state is
transformed into another pure coherent state which exhibits the
same minimal uncertainty. In fact, this observation confirms the
validity of using vacuum state for the initial calibration.

There exists, however, a stronger version of the Heisenberg
uncertainty relation which takes into account the purity of the
state, namely,
\begin{equation}
\label{purity-heisenberg} \sigma_{qq} \sigma_{pp} \ge \hbar^2
\Phi^2(\mu) / 4,
\end{equation}

\noindent which is also known as purity-dependent uncertainty
relation~\cite{dodonov-manko-1989}. Here, the purity-dependent
function $\Phi(\mu) = 2-\sqrt{2\mu-1}$ if $\frac{5}{9} \le \mu \le
1$ and $\Phi(\mu) \approx (4+\sqrt{16+9\mu^2}) / 9\mu \pm 4\%$
within the whole range $\mu \in (0,1]$. Employing the previously
found values of the purity (Sec.~\ref{section:purity}), the
inequality (\ref{purity-heisenberg}) transforms into $0.101 \pm
0.006 \ge 0.085 \pm 0.006$ for the SPACS, which is the first
direct experimental verification of formula
(\ref{purity-heisenberg}) within the accuracy $\sim 3\sigma$. This
result also encourages a feasible verification of two-mode
uncertainty relations~\cite{manko-ventriglia-2012} because the
corresponding methods of detecting two-mode states by a single
homodyne detector are already available~\cite{solimeno}.


\subsubsection{State-extended uncertainty relation}

Recently, Trifonov generalized uncertainty relations for a pair of
different states~\cite{trifonov-2000,trifonov-2002}, where the
variances of one state were connected with the variances of the
other by a series of so-called state-extended uncertainty
relation. One of such relations reads
\begin{equation}
\label{state-extended} \tfrac{1}{2} \big( \sigma_{qq}^{(1)}
\sigma_{pp}^{(2)} + \sigma_{qq}^{(2)} \sigma_{pp}^{(1)} \big) \ge
\tfrac{\hbar^2}{4}.
\end{equation}

\noindent We associate states ``1" and ``2" with the coherent
state and the SPACS, respectively. Using the experimental data,
the relation (\ref{state-extended}) takes the form $0.160 \pm
0.006
> 0.0625$, and thus is fulfilled with a great margin.
The great margin is due to the fact that ``1" is a coherent state
for which $\sigma_{qq}=\sigma_{pp}=\hbar/2$. This first
demonstration of state-extended uncertainty relation can encourage
its further applications to other states saturating it (e.g., some
squeezed states).


\subsection{Experimental check of entropic relations}


\subsubsection{Shannon entropy}

Given a wavefunction $\psi(q)$ of some pure state and the
wavefunction $\tilde{\psi}(p)$ of the same state in the momentum
representation, we are aware that they are not independent and are
related by the Fourier transform. In view of this, the narrower
the distribution $|\psi(q)|^2$ the wider $|\tilde{\psi}(p)|^2$ is
and vice versa. It means that the entropies $S_q = - \int
|\psi(q)|^2 \ln |\psi(q)|^2 dq $ and $S_p = - \int
|\tilde{\psi}(p)|^2 \ln |\tilde{\psi}(p)|^2 dp $ cannot take small
values simultaneously and turn out to satisfy the following
relation~\cite{bialynicki-birula-1975}:
\begin{equation}
\label{entropic-q-p} S_q + S_p \ge \ln (\pi \hbar) + 1,
\end{equation}

\noindent which is also valid in case of mixed states, with
$|\psi(q)|^2$ and $|\tilde{\psi}(p)|^2 $being replaced by the
marginal distributions $w(X,0)$ and $w(X,\frac{\pi}{2})$,
respectively. Since quadrature operators $\hat{X}_{\theta}$ and
$\hat{X}_{\theta+\pi/2}$ satisfy the same commutation relation as
$\hat{q}$ and $\hat{p}$ do, one can readily generalize
(\ref{entropic-q-p}) and
write~\cite{denicola,mmanko-2010,mmanko-2011}
\begin{equation}
\label{entropic-X}S(\theta) + S(\theta + \pi/2) \ge \ln (\pi
\hbar) + 1,
\end{equation}

\noindent where $S(\theta) \equiv S_{X_{\theta}}$ and the
right-hand side equals 1.45 if $\hbar = \tfrac{1}{2}$. The
inequality (\ref{entropic-X}) holds true for all LO phases
$\theta$. Considering $\theta$ as an additional independent
variable, one can now integrate (\ref{entropic-X}) over
$\theta\in[0,\pi]$. Taking into account that the theoretical
tomogram satisfies $w(X,\theta) = w(-X, \theta+\pi)$, we obtain
\begin{equation}
\label{entropic-theta} 2 \int_{0}^{\pi} S(\theta) \frac{d
\theta}{\pi} \ge \ln (\pi \hbar) + 1
\end{equation}

\noindent or, equivalently,
\begin{eqnarray}
\label{entropy-shannon-tomogram} H_{X,\theta} & \equiv & -
\int_{-\infty}^{+\infty} d X \int_{0}^{\pi}
\frac{d \theta}{\pi} ~ w(X,\theta) \ln w(X,\theta) \nonumber\\
& \ge & \tfrac{1}{2} \left[ \, \ln (\pi \hbar) + 1 \right],
\end{eqnarray}

\noindent where $H_{X,\theta}$ can be treated as the conventional
Shannon entropy of the probability distribution function
$w(X,\theta)$ of \emph{two} random variables $X\in
(-\infty,+\infty)$ and $\theta\in [0,\pi]$ such that
$\frac{1}{\pi} \int_{-\infty}^{+\infty} d X \int_{0}^{\pi} d
\theta ~ w(X,\theta) = 1$. A similar treatment of the quantum
homodyne tomography as an informationally complete positive
operator-valued measure on $[0,2\pi] \times \mathbb{R}$ is
presented in the paper~\cite{albini-devito-toigo}. From the
viewpoint of foundations of quantum mechanics, a quantum state is
defined by a fair probability distribution function $w(X,\theta)$
of two random variables (a point in the simplex of infinite
dimension) such that its entropy necessarily satisfies the
relation (\ref{entropy-shannon-tomogram}).

The evaluation of the integral (\ref{entropic-theta}) is performed
by a trapezoid method, i.e. $2 \int_{0}^{\pi} S(\theta) \frac{d
\theta}{\pi} \approx \sum_{j=1}^{N_{\theta}-1}
(\theta_{j+1}-\theta_j) [ S(\theta_j) + S(\theta_{j+1}) ]$. The
evaluation of $S(\theta_j)$, in its turn, is performed by
substituting the experimental binned histogram $h(X,\theta_j)$ for
$w(X,\theta)$.

The finite bin width $b$ is known to affect the right-hand side of
the relation (\ref{entropic-X}) (see~\cite{rudnicki} and
references therein). If we take the bin width $b=0.075$ and the
cutoff value of $X$ equal to 3, then the right-hand side of
(\ref{entropic-X}) is to be diminished by $0.03$ and equals $1.42$
for the choice $\hbar=\frac{1}{2}$. In fact, the allowance is
always negative and vanishes for larger cutoffs because the states
of our interest are localized quite close to center of the phase
space. Using quadratures $X_{\theta=0}$ and $X_{\theta=\pi/2}$, we
calculate the left-hand side of (\ref{entropic-q-p}) and the
result is $1.43 \pm 0.01$ for the coherent state and $1.65 \pm
0.03$ for the SPACS, where the errors are evaluated by comparing
the experimental values $S(\theta)$ and $S(\theta + \pi)$. The
coherent state saturates the boundary as it is predicted by the
theory~\cite{bialynicki-birula-1975}.

As to integral relation (\ref{entropic-theta}), the experimental
data yield the following quantities of the left hand side of
(\ref{entropic-theta}): $1.42 \pm 0.01$ for the coherent state and
$1.70 \pm 0.03$ for the SPACS, where the error bars comprise both
the error of calculation and the errors of the experimental data.


\subsubsection{R\'{e}nyi entropy}
\begin{figure}
\begin{center}
\includegraphics[width=8cm]{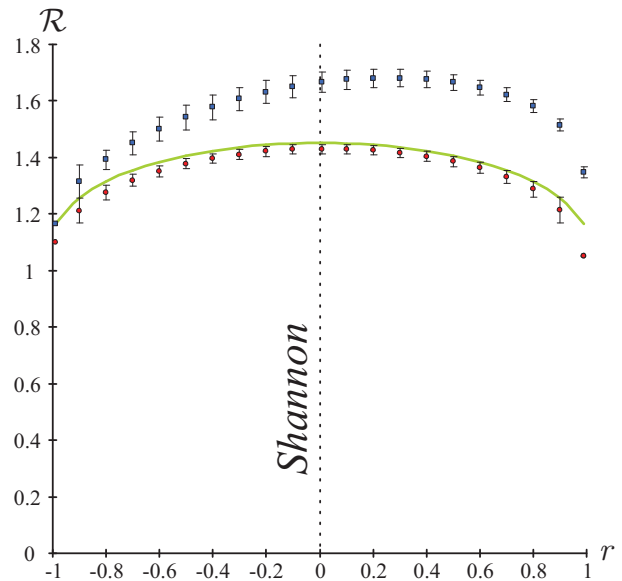}
\caption{\label{fig:renyi} (Color online) Experimental check of
uncertainty relations for the R\'{e}nyi entropy (\ref{renyi-r}).
Filled circles and squares correspond to experimental coherent and
SPAC states, respectively (for points without error bars the
corresponding errors are greater than 1). Solid line is a
theoretical bound. When $r\rightarrow 0$ the inequality
(\ref{renyi-r}) transforms into the inequality for the Shannon
entropy (\ref{entropic-q-p}).}
\end{center}
\end{figure}

The R\'{e}nyi entropy of the probability distribution $p(X)$ is
defined through $R_{\beta} [p(X)] = (1-\beta)^{-1} \ln \int
p^{\,\beta}(X) dX$ and represents nothing else but a
one-parametric family of entropy measures~\cite{renyi}. The
R\'{e}nyi entropy reduces to the Shannon entropy in the limit
$\beta \rightarrow 1$. For $\beta>1$ there exists a conjugate
parameter $\gamma$ such that $\beta^{-1}+\gamma^{-1}=2$. We can
put $\beta=(1-r)^{-1}$ and $\gamma=(1+r)^{-1}$, where $r\in(0,1)$.
An analog of the relation (\ref{entropic-q-p}) in terms of the
R\'{e}nyi entropy is $R_{\beta}[w(X,0)] +
R_{\gamma}[w(X,\frac{\pi}{2})] \ge \ln (\pi \hbar) - \frac{1}{2} [
(1-\beta)^{-1} \ln\beta + (1-\gamma)^{-1} \ln\gamma]$. In terms of
a single parameter $r$ this relation takes the
form~\cite{mmanko-2011}
\begin{eqnarray}
\label{renyi-r} && \!\!\!\!\! \mathcal{R}(r) \equiv \frac{1+r}{r}
\ln \left\{ \int [w(X,0)]^{(1+r)^{-1}} dX \right\}
\nonumber\\
&& \!\!\!\!\! - \frac{1-r}{r} \ln \left\{ \int
[w(X,\tfrac{\pi}{2})]^{(1-r)^{-1}} dX \right\} \nonumber\\
&& \!\!\!\!\! \ge \ln (\pi \hbar) + \tfrac{1}{2r} [ (1+r) \ln(1+r)
- (1-r) \ln(1-r)], \qquad
\end{eqnarray}

\noindent which remains true by replacing $w(X,0)$ and
$w(X,\frac{\pi}{2})$ by $w(X,\theta)$ and $w(X,\frac{\pi}{2} +
\theta)$, respectively. For the sake of simplicity, we concentrate
on the experimental check of inequality (\ref{renyi-r}) for
$r\in(-1,1)$. The results are presented in Fig.~\ref{fig:renyi},
where the points correspond to the left-hand side of
(\ref{renyi-r}) calculated via the experimental histograms. As
above, the coherent state saturates the boundary (within the
experimental errors) and the experimentally determined values are
symmetrical with respect to change $r \rightarrow -r$ because the
position and momentum are identically distributed. This does not
take place for the SPACS and such an asymmetry is readily seen.

\section{\label{section:summary} Summary}
In this paper, we have considered a relatively simple but
extremely powerful experimental apparatus to measure quantum
states of light -- homodyne detector. Our main idea was to use
measurable quantities (histograms) to reveal as much information
about light states as possible. First, the measured histograms
enabled us to estimate the optical tomogram, i.e. the quantum
state itself. We developed a method for choosing an optimal bin
width, which ensures statistical confidence and prevents from
undersampling at the same time. Second, we managed to accomplish a
quantitative analysis of the accuracy of estimated tomograms by
using the peculiar property of fair tomograms $w(X,\theta) =
w(-X,\theta + \pi)$. Distinction of our approach is that the
evaluated errors comprise both statistical and systematical
errors. Moreover, the detailed analysis can also reveal probable
sources of systematical errors such as imprecision of the LO phase
control, which is hardly possible to detect by other methods. Even
if the systematic error cannot be got rid of, one can use an
original collection of experimental data to postselect those data
which exhibit the least systematic error. Third, we used the
measurable quantities (histograms) to calculate the
characteristics of the state directly. For instance, the purity
and its error are naturally calculated on the basis of measured
experimental data without any time-consuming state-reconstruction
procedure with controversial error estimation. Our data result in
the relative error about several percent (1$\div$5\%) for almost
all state characteristics (when their theoretical values do not
vanish). Last but not least, the operational use of the data
allowed us to check for the first time the fundamental properties
of quantum objects such as the (purity-dependent) uncertainty
relations for position and momentum as well as their entropic
analogs.

To conclude, we believe that the developed methods will contribute
to achieving a higher precision of optical homodyne detection and
encourage the operational use of experimental data, which can turn
out to be crucial for the analysis of multimode quantum states.


\begin{acknowledgments}
The authors thank the anonymous referee for insightful and
constructive comments. M.B. and A.Z. acknowledge support of Ente
Cassa di Risparmio di Firenze, Regione Toscana under project
CTOTUS, EU under ERA-NET CHIST-ERA project QSCALE, and MIUR, under
contract FIRB RBFR10M3SB. A.S.C. acknowledges total financial
support from the Funda\c{c}\~ao de Amparo \`a Pesquisa do Estado
S\~ao Paulo (FAPESP). S.N.F. and V.I.M. thank the Russian
Foundation for Basic Research for partial support under projects
10-02-00312 and 11-02-00456 and the Ministry of Education and
Science of the Russian Federation for partial support under
project no. 2.1759.2011. S.N.F. acknowledges the support of the
Dynasty Foundation and the Ministry of Education and Science of
the Russian Federation (projects 2.1.1/5909, $\Pi$558, and
14.740.11.1257).
\end{acknowledgments}


\end{document}